\documentclass[journal]{IEEEtran}
\usepackage[colorlinks,
linkcolor=red,
anchorcolor=green,
citecolor=blue
]{hyperref} 
\usepackage{amsmath,amssymb}
\usepackage{latexsym}
\usepackage{CJK}
\usepackage{cite}

\usepackage{color, soul}

\usepackage{graphicx}
\usepackage{epstopdf}
\usepackage{epsfig}
\usepackage{subfigure} 

\usepackage{verbatim}  

\usepackage{mathrsfs}	
\usepackage{algorithm} 
\usepackage{algorithmic} 
\usepackage{booktabs}
\usepackage{textcomp}
\usepackage{multirow}
\usepackage{lettrine}

\usepackage{mathrsfs}
\usepackage{amsmath}
\usepackage{amssymb}

\usepackage{algorithm}
\usepackage{algorithmic}
\usepackage{extarrows}

\usepackage{amsthm} 

\newtheorem{prop}{Proposition}
\newtheorem{defn}{Definition}
\newtheorem{rem}{Remark}

{ \bgroup
  \addtolength\abovedisplayshortskip{#1}
  \addtolength\abovedisplayskip{#1}
  \addtolength\belowdisplayshortskip{#1}
  \addtolength\belowdisplayskip{#1}}
{\egroup\ignorespacesafterend}

\usepackage{subeqnarray}
\usepackage{cases}
\usepackage{mathtools} 
\usepackage{bm}

\ifCLASSINFOpdf
\else
\fi
\begin{document}
%
\title{
Information Measure Similarity Theory: Message Importance Measure via Shannon Entropy}
\author{\IEEEauthorblockN{Rui She,
Shanyun Liu, and
Pingyi Fan
\\}

\thanks{
This work was supported by the National Natural Science Foundation of China (NSFC) No. 61771283.


R. She, S. Liu, and P. Fan are with the Department of Electronic Engineering, Tsinghua University, Beijing, 100084, China (e-mail: sher15@mails.tsinghua.edu.cn; liushany16@mails.tsinghua.edu.cn; fpy@tsinghua.edu.cn).
%
%
}
}
\maketitle

\begin{abstract}
Rare events attract more attention and interests in many scenarios of big data such as anomaly detection and security systems.
To characterize the rare events importance from probabilistic perspective,
the message importance measure (MIM) is proposed as a kind of semantics analysis tool.
Similar to Shannon entropy, the MIM has its special functional on information processing, in which the parameter $\varpi$ of MIM plays a vital role.
Actually, the parameter $\varpi$ dominates the properties of MIM, based on which the MIM has three work regions where the corresponding parameters satisfy $ 0 \le \varpi \le 2/\max\{p(x_i)\}$, $\varpi  > 2/\max\{p(x_i)\}$ and $\varpi < 0$ respectively.
Furthermore, in the case $ 0 \le \varpi \le 2/\max\{p(x_i)\}$, there are some similarity between the MIM and Shannon entropy in the information compression and transmission, which provide a new viewpoint for information theory.
This paper first constructs a system model with message importance measure and proposes the message importance loss to enrich the information processing strategies.
Moreover, we propose the message importance loss capacity to measure the information importance harvest in a transmission.
Furthermore, the message importance distortion function is presented to give an upper bound of information compression based on message importance measure.
Additionally, the bitrate transmission constrained by the message importance loss is investigated to broaden the scope for Shannon information theory.
\end{abstract}

\begin{IEEEkeywords}
Message importance measure, information theory, big data analytics and processing, message transmission and compression
\end{IEEEkeywords}

\IEEEpeerreviewmaketitle

\section{Introduction}
\IEEEPARstart{I}{n} recent years, massive data has attracted much attention in various realistic scenarios, which is called the ``big data era''.
In this context, it is a key point that how to deal with the observed data and dig the hidden valuable information out of the collected data \cite{Deep-learning,
DNN-filter-bank,Text-independent-speaker}.
To do so, a series of efficient technologies have been put forward such as learning tasks, computer vision, image recognition and neural networking \cite{Big-Data-Deep-Learning,Toward-scalable-systems-for-big-data-analytics,Machine-learning-with-big-data,
Information-security-in-big-data}.

In fact, there still exist many challenges for big data analytics and processing such as distributed data acquisition, huge-scale data storage and transmission, and decision-making based on individualized requirements.
Facing these obstacles in big data, it is promising to combine information theory and probabilistic statistics with events semantics to deal with massive information.
To some degree, more attention is paid to rare events than those with large probability.
Due to the fact that small probability events containing semantic importance may be hidden in big data \cite{Efficient-algorithms-for-mining-outliers-from-large-data-sets,
Improved-principal-component,An-improved-methodology,Nonlinear-Gaussian-belief,
Principal-components-selection}, it is significant to process rare events or the minority in numerous applications such as outliers detection in the Internet of Things (IoTs), smart cities and autonomous driving \cite{Blockchains-and-Smart-Contracts,
Design-and-realization,A-Survey-on-Internet,Internet-of-things-and-big-data,
Internet-of-Things-for-Smart-Cities,An-anomaly-detection-in-smart-cities,
Detecting-unexpected-obstacles-for-self-driving-cars,
Lane-following-and-obstacle-detection-techniques,An-improved-lane-departure-method}.
Therefore, the rare events have special values in the data mining and processing based on semantics analysis of message importance.

In order to characterize rare events importance in big data, a new information measure named message importance measure (MIM) is presented to generalize Shannon information theory \cite{Message-Importance-Measure-and-Its-Application,Focusing-on-a-Probability-Element}.
For convenience of calculation, an exponential expression of MIM is obtained as follows.
\begin{defn}\label{defn:MIM}
For a discrete distribution $P(X)$=$\{p(x_1)$, $p(x_2)$, ...,$ p(x_n)\}$, the exponential expression of message importance measure (MIM) is given by
\begin{equation}\label{MIM_discrete1}
 \begin{aligned}
    L(\varpi, X)
    & = \sum\limits_{x_i} p(x_i) e^{\varpi\{1-p(x_i)\}},\\
 \end{aligned}
\end{equation}
where the adjustable parameter $\varpi$ is nonnegative and $p(x_i) e^{\varpi\{1-p(x_i)\}}$ is viewed as the self-scoring value of event $i$ to measure its message importance.
\end{defn}

Actually, from the perspective of generalized Fadeev's postulates, the MIM is viewed as a rational information measure similar to Shannon entropy and Renyi entropy.
In particular,
a postulate for the MIM weaker than that for Shannon entropy and Renyi entropy is given by
\begin{equation}
    F(PQ) \le F(P) + F(Q),
\end{equation}
while $F(PQ) = F(P) + F(Q)$ is satisfied in Shannon entropy and Renyi entropy \cite{On-measures-of-entropy-and-information}, where $P$ and $Q$ are two independent random distributions and $F(\cdot)$ denotes a kind of information measure.

\subsection{ The importance coefficient $\varpi$ in MIM}\ \par
In general, the parameter $\varpi$ viewed as importance coefficient, has a great impact on the MIM. Actually, different parameter $\varpi$ can lead to different properties and performances for this information measure.
In particular, to measure a distribution $P(X)=\{p(x_1),p(x_2),...,p(x_n)\}$, there are three kinds of work regions of MIM which can be classified by the parameters, whose details are discussed as follows.

\textit{i)} If the parameter satisfies $ 0 \le \varpi \le 2/\max\{p(x_i)\}$, the convexity of MIM is similar to Shannon entropy and Renyi entropy. Actually, these three information measures all have maximum value property and can emphasize small probability elements of the distribution $P(X)$ in some degree. It is notable that the MIM in this work region focuses on the typical sets rather than atypical sets and the uniform distribution reaches the maximum value.
In brief, the MIM in this work region can be regarded as the same message measure as Shannon entropy and Renyi entropy to deal with the problems of information theory such as data compression, storage and transmission.

\textit{ii)} If we have $\varpi  > 2/\max\{p(x_i)\}$, the small probability elements will be the dominant factor for MIM to measure a distribution. That is, the small probability events can be highlighted more in this work region of MIM than those in the first one. Moreover, in this work region MIM can pay more attention to atypical sets, which can be viewed as a magnifier for rare events.
In fact, this property corresponds to some common scenarios where anomalies catch more eyes such as anomalous detection and alarm. In this case, some problems (including communication and probabilistic events processing) can be rehandled from the perspective of rare events importance. Particularly, the compression encoding and maximum entropy rate transmission are proposed based on the non-parametric MIM \cite{Storage-Code-Design-and-Transmission}, as well as, the distribution goodness-of-fit approach is also presented by use of differential MIM \cite{Differential-message-importance-measure}.

\textit{iii)} If the MIM has the parameter $\varpi < 0$, the large probability elements will be the main part contributing to the value of this information measure. In other words, the normal events attract more attention in this work region of MIM than rare events. In practice, there are many applications where regular events are popular such as filter systems and data cleaning.

As a matter of fact, by selecting the parameter $\varpi$ properly, we can exploit the MIM to solve several problems in different scenarios. The importance coefficient facilitates more flexibility of MIM in applications beyond Shannon entropy and Renyi entropy.

To focus on a concrete object, in this paper we mainly investigate the first kind of MIM ($0 \le \varpi \le 2/\max\{p(x_i)\}$) and intend to dig out some novelties related to this information measure.

\subsection{Message importance measure similar to Shannon entropy}\ \par
From the perspective of information flow processing, there are some distortions for probabilistic events during the information compression and transmission.
However, rare events with much message importance require higher reliability than those with large probability.
In this regard, Shannon information theory can be amended for the big data processing based on the message importance of rare events.
Thus, some information measures with respect to message importance are investigated to extend the range of Shannon information theory \cite{Information-theoretic-measures-for-anomaly-detection,
An-information-theoretic-approach-to-detection-of-minority-subsets-in-database,
The-large-deviation,A-measure-of-the-concentration,A-large-deviations-approach}.
In this case, the MIM is considered as a kind of promising information measure supplemented to Shannon entropy and Renyi entropy.

In some degree, when the parameter $\varpi$ satisfies $ 0 \le \varpi \le 2/\max\{p(x_i)\}$, the MIM is similar to Shannon entropy from the perspective of expression and properties. The exponential operator of MIM is a substitute for logarithm operator of Shannon entropy. This implies that small probability elements are amplified more in the MIM than those in Shannon entropy. However, as a kind of tool to measure probability distribution, the MIM with parameter
$ 0 \le \varpi \le 2/\max\{p(x_i)\}$ has the same concavity and monotonicity as Shannon entropy, which can characterize the information otherness for different variables.

In addition, similar to Shannon conditional entropy, a conditional message importance measure for two distributions is proposed to process conditional probability.
\begin{defn}\label{defn:CMIM}
For the two discrete probability $P(X)$=$\{p(x_1)$, $p(x_2)$, ...,$ p(x_n)\}$ and $P(Y)$=$\{p(y_1)$, $p(y_2)$, ...,$ p(y_n)\}$, the conditional message importance measure (CMIM) is given by
\begin{equation}\label{CMIM_discrete1}
 \begin{aligned}
    L(\varpi, Y|X)
    & = \sum\limits_{x_i} p(x_i) \sum\limits_{y_i} p(y_j|x_i) e^{\varpi\{1-p(y_j|x_i)\}},\\
 \end{aligned}
\end{equation}
where $p(y_j|x_i)$ denotes the conditional probability between $y_j$ and $x_i$.
The component
$p(y_j|x_i) e^{\varpi\{1-p(y_j|x_i)\}}$ is similar to
self-scoring value. 
Therefore, the CMIM can be considered as a system invariant which indicates the average total self-scoring value for a information transfer process.
\end{defn}


In fact, due to the similarity between the MIM and Shannon entropy, they may have analogous performance on information processing including data compression and transmission.
To this end, a new information measure theory based on the MIM is discussed in this paper.

\subsection{Organization}
The rest of this paper is discussed as follows.
In Section II, a system model involved with message importance is constructed to help analyze the data compression and transmission in big data.
In Section III, we propose a kind of message transfer capacity to investigate the message importance loss in the transmission.
In Section IV, message importance distortion function is introduced and its properties are also presented to give some details.
In Section V, we discuss the bitrate transmission constrained by message importance to widen the horizon for the Shannon theory.
In Section VI, some numerical results are presented to validate propositions and the analyses in theory.
Finally, we conclude this paper in Section VII.

\section{System Model with message importance}
\begin{figure*}
\centering
\includegraphics[width=5.6in]{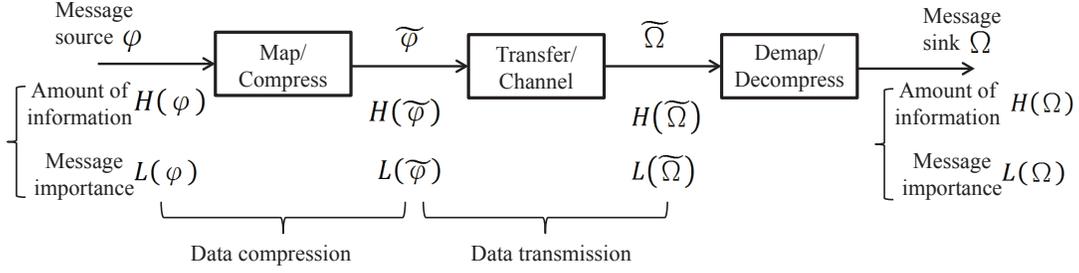}
\caption{Information processing system model. }
\label{fig:system model}
\end{figure*}

Consider an information processing system model shown in Fig. \ref{fig:system model}, in which the information transfer process is discussed as follows.
At first, a message source $\varphi$ follows a distribution $P=(p_1,p_2,...,p_n)$ whose support set is $\{\varphi_1,\varphi_2,...,\varphi_n\}$ corresponding to the events types.
Then, the message $\varphi$ is encoded or compressed into the variable $\widetilde \varphi$ following the distribution $P_{\widetilde \varphi}=(p_{\widetilde \varphi_1},p_{\widetilde \varphi_2},...,p_{\widetilde \varphi_n})$ whose alphabet is $\{\varphi_1,\varphi_2,...,\varphi_n\}$.
In this case, the sample sequence $\widetilde \varphi^N= \{\widetilde \varphi_1,\widetilde \varphi_2,...,\widetilde \varphi_N\}$ drawn from $\widetilde \varphi$ satisfies the asymptotic equipartition property (AEP).
After the information transfer process denoted by matrix $p(\widetilde \Omega_j|\widetilde \varphi_i)$, the received message $\widetilde \Omega$ originating from $\widetilde \varphi$ is observed as a random sequence $\widetilde \Omega^N$, where
the distribution of $\widetilde \Omega $ is $P_{\widetilde \Omega}=(p_{\widetilde \Omega_1},p_{\widetilde \Omega_2},...,p_{\widetilde \Omega_n})$ whose alphabet is $\{\widetilde \Omega_1,\widetilde \Omega_2,...,\widetilde \Omega_n\}$.
At last, the receiver recovers the original message $\varphi$ by decoding ${\Omega}=g(\widetilde \Omega^N)$ where $g(\cdot)$ denotes the decoding function and $\Omega$ is the recovered message with the alphabet $\{\Omega_1,\Omega_2,...,\Omega_n\}$.

Actually, different from the mathematically probabilistic characterization of a traditional telecommunication system, this paper discusses the information processing from two perspectives of information, namely the amount of information $H(\cdot)$ and message importance ${L}(\cdot)$.
In particular, from the viewpoint of generalized information theory, a two-layer framework is considered to understand this model, where the first layer is based on the amount of information characterized by Shannon entropy, while the second layer reposes on message importance of rare events.
Due to the fact that the former is discussed pretty entirely, we mainly investigate the latter in the paper.

In addition, considering the source-channel separation theorem \cite{Elements-of-information-theory-2nd-edition}, the above information processing model consists of two problems, namely data compression and data transmission.
On one hand, the \textit{data compression} of the system can be achieved by using classical source coding strategies to reduce more redundancy, in which the information loss is described by $H(\varphi)-H(\varphi|\widetilde \varphi)$ under the information transfer matrix $p(\widetilde \varphi|\varphi)$.
Similarly, from the perspective of message importance, the data can be further compressed by discarding worthless messages, where the message importance loss can be characterized by
$L(\varphi)-L(\varphi| \widetilde \varphi)$.
On the other hand, the \textit{data transmission} is discussed to obtain the upper bound of the mutual information $H(\widetilde \varphi)- H(\widetilde \varphi|\widetilde \Omega)$, namely the information capacity. In a similar way, $L(\widetilde \varphi)- L(\widetilde \varphi|\widetilde \Omega)$ means the income of message importance in the transmission.

In essence, it is apparent that the data compression and transmission are both considered as an information transfer processes $\{X,p(y|x),Y\}$, and they can be characterized by the difference between $\{X\}$ and $\{X|Y\}$.
In order to facilitate the analysis of the above model, the message importance loss is introduced as follows.

\begin{defn}\label{defn:Loss}
For two discrete probability $P(X)$=$\{p(x_1)$, $p(x_2)$, ...,$ p(x_n)\}$ and $P(Y)$=$\{p(y_1)$, $p(y_2)$, ...,$ p(y_n)\}$, the message importance loss based on MIM and CMIM is given by
\begin{equation}\label{Loss}
 \begin{aligned}
    \Phi_{\varpi} (X||Y)
    & = L(\varpi, X) - L(\varpi, X|Y),\\
 \end{aligned}
\end{equation}
where $L(\varpi, X)$ and $L(\varpi, X|Y)$ are given by the Definition \ref{defn:MIM} and \ref{defn:CMIM}.
\end{defn}

In fact, according to the intrinsic relationship between $L(\varpi, X)$ and $L(\varpi, X|Y)$, it is readily seen that
\begin{equation}
 \begin{aligned}
    \Phi_{\varpi} (X||Y)\ge 0,
 \end{aligned}
\end{equation}
where $0<\varpi \le 2\le 2/\max\{{p(x_i|y_j)}\}$.
\begin{IEEEproof}
Considering a function $f(x)=xe^{\varpi(1-x)}$ ($0\le x \le 1$ and $0< \varpi$), it is easy to have
$\frac{\partial^2 f(x)}{\partial x} = -\varpi e^{\varpi(1-x)}(2-\varpi x)$,
which implies if $\varpi \le 2\le {2}/{x}$, the function $f(x)$ is concave.

In the light of Jensen's inequality, if $0<\varpi \le 2\le 2/\max\{{p(x_i|y_j)}\}$ is satisfied, it is not difficult to see
\begin{equation}
 \begin{aligned}
    & L(\varpi, X) \\
    & = \sum_{x_i}p(x_i)e^{\varpi(1-p(x_i))}\\
    & = \sum_{x_i}\{\sum_{y_j}p(y_j)p(x_i|p_j)\}e^{\varpi(1-\{\sum_{y_j}p(y_j)p(x_i|p_j)\})}\\
    & \ge \sum_{y_j}p(y_j)\sum_{x_i}\{p(x_i|p_j)e^{\varpi(1-p(x_i|p_j))}\}
    = L(\varpi,X|Y),
 \end{aligned}
\end{equation}
which testifies the nonnegativity of $\Phi_{\varpi} (X||Y)$.
\end{IEEEproof}

\section{Message importance loss in transmission }
In this section,
we will introduce the CMIM to characterize the information transfer processing.
To do so, we define a kind of message transfer capacity measured by the CMIM as follows.
\begin{defn}\label{defn:C_L}
Assume that there exists
an information transfer process as
\begin{equation}\label{eq.relation}
 \begin{aligned}
    \{ X, p(y|x), Y\},
 \end{aligned}
\end{equation}
where the $p(y|x)$ denotes a probability distribution matrix describing the information transfer from the variable $X$ to $Y$.
We define the message importance loss capacity (MILC) as
\begin{equation}\label{eq.D_channel}
   \begin{aligned}
    C
    & = \max\limits_{p(x)} \{ \Phi_{\varpi} (X||Y) \}\\
    & = \max\limits_{p(x)} \{ L(\varpi, X) - L(\varpi, X|Y)\},\\
   \end{aligned}
\end{equation}
where $L(\varpi, X) = \sum_{x_i} p(x_i) e^{\varpi\{1-p(x_i)\}}$, $p(y_j) = \sum_{x_i} p(x_i)p(y_j|x_i)$,
$p(x_i|y_j) = \frac{p(x_i)p(y_j|x_i)}{p(y_j) } $, $L(\varpi, X|Y)$ is defined by Eq. (\ref{CMIM_discrete1}), and $\varpi < 2 \le 2/\max{\{ p(x_i) \} }$.
\end{defn}

%
In order to have an insight into the applications of MILC, some specific information transfer scenarios are discussed as follows.

\subsection{Binary symmetric matrix}
Consider the binary symmetric information transfer matrix, where the original variables are complemented with the transfer probability which can be seen in the following proposition.

\begin{prop}\label{prop.symmetric}
Assume that there exists an information transfer process $\{ X, p(y|x), Y\}$,
where the information transfer matrix is
\begin{equation}\label{eq.symmetric_channel}
 \begin{aligned}
   p(y|x) = \left [
   \begin{matrix}
    1-\beta_{s} & \beta_{s} \\
    \beta_{s} & 1-\beta_{s}
   \end{matrix}
   \right ],
 \end{aligned}
\end{equation}
which indicates that $X$ and $Y$ both follow binary distributions.
In that case, We have
\begin{equation}\label{eq.C_binary}
\begin{aligned}
    C(\varpi, \beta_{s}) = e^{ \frac{\varpi}{2} } - L(\varpi, \beta_{s}),
\end{aligned}
\end{equation}
where $L(\varpi, \beta_{s})= \beta_{s}  e^{\varpi(1-\beta_{s}) } + (1-\beta_{s} )e^{\varpi \beta_{s}} $ ($0\le \beta_{s} \le 1$) and $\varpi < 2 \le 2/\max{\{ p(x_i) \} }$.
\end{prop}

\begin{IEEEproof}
Assume that the distribution of variable $X$ is a binary distribution $(p, 1-p)$. According to Eq. (\ref{eq.symmetric_channel}) and Bayes' theorem (namely, $p(x|y) = \frac{p(x)p(y|x)}{p(y)}$),
it is not difficult to see that
\begin{equation}\label{eq.symmetric_X_Y}
 \begin{aligned}
   p(x|y) = \left [
   \begin{matrix}
    \frac{p(1-\beta_{s})}{p(1-\beta_{s})+(1-p)\beta_{s}} & \frac{(1-p)\beta_{s}}{p(1-\beta_{s})+(1-p)\beta_{s}} \\
    \frac{p\beta_{s}}{p\beta_{s}+(1-p)(1-\beta_{s})} & \frac{(1-p)(1-\beta_{s})}{p\beta_{s}+(1-p)(1-\beta_{s})}
   \end{matrix}
   \right ].
 \end{aligned}
\end{equation}

Furthermore, in accordance with Eq. (\ref{CMIM_discrete1}) and Eq. (\ref{eq.D_channel}), we have
\begin{equation}
\begin{aligned}
     & C(\varpi, \beta_{s} ) = \max\limits_{p} \{C(p, \varpi, \beta_{s} )\}\\
     & = \max\limits_{p} \Big\{ L(\varpi, p)
      - \big\{ p(1-\beta_{s})e^{\frac{\varpi (1-p)\beta_{s}}{p(1-\beta_{s})+(1-p)\beta_{s}}} \\
     & + (1-p)\beta_{s} e^{\frac{\varpi p(1-\beta_{s})}{p(1-\beta_{s})+(1-p)\beta_{s}}}
     + p\beta_{s} e^{\frac{\varpi(1-p)(1-\beta_{s})}{p\beta_{s}+(1-p)(1-\beta_{s})}} \\
     & + (1-p)(1-\beta_{s}) e^{\frac{\varpi p\beta_{s}}{p\beta_{s}+(1-p)(1-\beta_{s})}}
     \big\} \Big\},
\end{aligned}
\end{equation}
where $L(\varpi, p)= p  e^{\varpi(1-p) } + (1-p )e^{\varpi p} $ ($0< p <1$).
Then, it is readily seen that
\begin{equation}\label{eq.dCbinary}
\begin{aligned}
     & \frac{\partial C(p,\varpi, \beta_{s} )}{\partial p} \\
     & = (1-\varpi p)e^{\varpi(1-p)} + [(1-p)\varpi-1]e^{\varpi p} \\
     & - \bigg\{ (1-\beta_{s})\Big\{ 1- \frac{\varpi p(1-\beta_{s})\beta_{s}}{[p(1-\beta_{s})+(1-p)\beta_{s}]^2} \Big\}e^{\frac{\varpi(1-p)\beta}{p(1-\beta)+(1-p)\beta}}\\
     & + (1-\beta_{s})\Big\{ \frac{\varpi(1-p)\beta_{s}(1-\beta_{s})}{[p\beta_{s}+(1-p)(1-\beta_{s})]^2} -1\Big\}e^{\frac{\varpi p\beta_{s}}{p\beta_{s}+(1-p)(1-\beta_{s})}}\\
     & + \beta_{s} \Big\{ \frac{\varpi(1-p)\beta_{s}(1-\beta_{s})}{[p(1-\beta_{s})+(1-p)\beta_{s}]^2}-1 \Big\}e^{\frac{\varpi p(1-\beta_{s})}{p(1-\beta_{s})+(1-p)\beta_{s}}}\\
     & + \beta_{s} \Big\{ 1- \frac{\varpi p(1-\beta_{s})\beta_{s}}{[p\beta_{s} + (1-p)(1-\beta_{s})]^2} \Big\} e^{\frac{\varpi(1-p)(1-\beta_{s})}{p\beta_{s} +(1-p)(1-\beta_{s})}}\bigg\}. \\
\end{aligned}
\end{equation}

In the light of the positive for $\frac{\partial C(p, \beta_{s} )}{\partial p}$ in $\{p| p \in (0,1/2)\}$ and the negativity in $\{ p| p \in (1/2,1)\}$ (if $\beta_{s} \ne 1/2$), it is apparent that $p=1/2$ is the only solution for $\frac{\partial C(p, \beta_{s} )}{\partial p} =0 $.
That is, if $\beta_{s} \ne 1/2$, the extreme value is indeed the maximum value of $C(p, \varpi, \beta_{s} )$ when $p=1/2$.
Similarly, if $\beta_{s} = 1/2$, the solution $p= 1/2$ also results in the same conclusion.
Therefore, by substituting $p=1/2$ into $C(p, \varpi, \beta_{s} )$, the proposition is testified.
\end{IEEEproof}

\begin{rem}
According to Proposition \ref{prop.symmetric}, on one hand, when $\beta_{s}=1/2$, that is, the information transfer process is just random, we will gain the lower bound of the MILC namely $C(\beta_{s})=0$. On the other hand, when $\beta_{s}=0$, namely there is a certain information transfer process, we will have the maximum MILC.
As for the distribution selection for the variable $X$, the uniform distribution is preferred to gain the capacity.
\end{rem}

\subsection{Binary erasure matrix}
The binary erasure information transfer matrix is similar to the binary symmetric one, however, in the former a part of information is lost rather than corrupted.
The MILC of this kind of information transfer matrix is discussed as follows.

\begin{prop}\label{prop.error_symmetric}
Consider an information transfer process $\{ X, p(y|x), Y\}$, in which the information transfer matrix is described as
\begin{equation}
 \begin{aligned}
   p(y|x) = \left [
   \begin{matrix}
    1-\beta_{e} & 0 &\beta_{e} \\
    0 & 1-\beta_{e} &\beta_{e}
   \end{matrix}
   \right ],
 \end{aligned}
\end{equation}
which indicates that $X$ follows the binary distribution and $Y$ follows the $3$-ary distribution.
Then, we have
\begin{equation}\label{eq.Cerasure}
\begin{aligned}
     C(\varpi, \beta_{e})
     & = (1-\beta_{e}) \{ e^{\frac{\varpi}{2}} - 1\},
\end{aligned}
\end{equation}
where $0\le \beta_{e} \le 1$ and $0<\varpi < 2 \le 2/\max{\{ p(x_i) \} }$.
\end{prop}

\begin{IEEEproof}
Assume the distribution of variable $X$ is $(p, 1-p)$. As well, according to the binary erasure matrix and Bayes theorem, we have the transmission matrix conditioned by the variable $Y$ as follows
\begin{equation}\label{eq.erasure_x_y}
\begin{aligned}
    p(x|y) = \left [
    \begin{matrix}
        1 & 0 \\
        0 & 1 \\
        p & 1-p
    \end{matrix}
    \right ].
\end{aligned}
\end{equation}
Then, it is not difficult to have
\begin{equation}\label{eq.erasure_Lx_y}
\begin{aligned}
      L(\varpi, X|Y)
     & =  \beta_{e} p e^{\varpi(1-p)} + \beta_{e} (1-p) e^{\varpi p} + 1-\beta_{e}.   \\
\end{aligned}
\end{equation}
Furthermore, it is readily seen that
\begin{equation}\label{eq.erasure_C}
\begin{aligned}
     & C(p, \varpi,\beta_{e})\\
     & = \max\limits_{p} \Big\{ L(\varpi, p)
     - \big\{  \beta_{e} p e^{\varpi(1-p)}
     + \beta_{e} (1-p) e^{\varpi p} + 1-\beta_{e} \big\}  \Big\} \\
     & = (1-\beta_{e})\big\{\max\limits_{p}\{L(\varpi, p)\} -1\big\},\\
\end{aligned}
\end{equation}
where $L(\varpi, p)= p e^{\varpi(1-p) } + (1-p )e^{\varpi p}$.
Moreover, we have the solution $p= 1/2$ leads to $\frac{\partial L(\varpi, p)}{\partial p}=0$
and the corresponding second derivative is
\begin{equation}
\begin{aligned}
     & \frac{\partial^2 L(\varpi, p)}{\partial p^2}\\
     & = e^{\varpi(1-p)}(\varpi p-2)\varpi + e^{\varpi p}[(1-p)\varpi-2]\varpi \\
     & < 0,
\end{aligned}
\end{equation}
which is resulted from the condition $0<\varpi < 2 \le 2/\max{\{ p(x_i) \} }$.

Therefore, it is readily seen that in the case $p=1/2$, the capacity $C(p,\varpi, \beta_{e}) $ reaches the maximum value, which testifies this proposition.
\end{IEEEproof}

\begin{rem}
Proposition \ref{prop.error_symmetric} indicates that in the case $\beta_{e}=1$, the lower bound of the capacity is obtained, that is $C(\beta_{e})=0$. However, if a certain information transfer process is satisfied namely $\beta_{e}=0$, we will have the maximum MILC.
Similar to the Proposition \ref{prop.symmetric}, the uniform distribution is selected in practice to reach the capacity.
\end{rem}

\subsection{Strongly symmetric backward matrix}
As for strongly symmetric backward matrix, it is viewed as a special example of information transmission.
The discussion for the message transfer capacity in this case is similar to that in the symmetric matrix, whose details are given as follows.
\begin{prop}\label{prop_3}
For an information transmission from the source $X$ to the sink $Y$, assume that there exists a strongly symmetric backward matrix as follows
\begin{equation}\label{eq.strongly_symmetric_matrix}
 \begin{aligned}
   p(x|y) = \left [
   \begin{matrix}
    1-\beta_{k} & \frac{\beta_{k}}{K-1} &...& \frac{\beta_{k}}{K-1} \\
    \frac{\beta_{k}}{K-1} & 1-\beta_{k} & ...& \frac{\beta_{k}}{K-1} \\
    ...& ... & ... & ...\\
    \frac{\beta_{k}}{K-1} &...& \frac{\beta_{k}}{K-1} &  1-\beta_{k}
   \end{matrix}
   \right ],
 \end{aligned}
\end{equation}
which indicates that $X$ and $Y$ both obey $K$-ary distribution.
We have
\begin{equation}\label{eq.C_strongly}
\begin{aligned}
    C(\varpi, \beta_{k}) = e^{\frac{\varpi(K-1)}{K}}- \{ (1-\beta_{k})e^{\varpi \beta_{k} } + \beta_{k} e^{\varpi(1-\frac{\beta_{k}}{K-1}) } \},
\end{aligned}
\end{equation}
where the parameter $0\le \beta_{k} \le 1$, $K\ge 2$ and $0<\varpi < 2 \le 2/\max{\{ p(x_i) \} }$.
\end{prop}

\begin{IEEEproof}
For given $K$-ary variables $X$ and $Y$ whose distribution are $\{p(x_1),p(x_2),...,p(x_K) \}$ and $\{p(y_1),p(y_2),...,p(y_K) \}$ respectively, we can use the strongly symmetric backward matrix to obtain the relationship between the two variables as follows
\begin{equation}\label{eq.p_x}
\begin{aligned}
    p(x_i) & = (1-\beta_{k})p(y_i) + \frac{\beta_{k}}{K-1}[1- p(y_i)], (i =1, 2, ..., K)\\
\end{aligned}
\end{equation}
which implies
$p(x_i)$ is a one-to-one onto function for $p(y_i)$.

In accordance with Definition \ref{defn:CMIM}, it is readily to see that
\begin{equation}
\begin{aligned}
    &  L(\varpi, X|Y)\\
    & = \sum\limits_{x_i} \sum\limits_{y_j} p(y_j) p(x_i|y_j) e^{\varpi(1-p(x_i|y_j))}\\
    & = \sum\limits_{y_j} p(y_j) \big\{ (1-\beta_{k})e^{\varpi \beta_{k}} + \beta_{k} e^{\varpi(1-\frac{\beta_{k}}{K-1})} \big\}\\
    & = (1-\beta_{k})e^{\varpi \beta_{k}} + \beta_{k} e^{\varpi(1-\frac{\beta_{k}}{K-1})}.
\end{aligned}
\end{equation}

Moreover, by virtue of the definition of MILC in Eq. (\ref{eq.D_channel}), it is easy to see that
\begin{equation}\label{eq.C_derivative}
\begin{aligned}
     & C(\varpi, \beta_{k})\\
     & = \max\limits_{p(x)} \{L(\varpi, X)\}
     - [\ (1-\beta_{k})e^{\varpi \beta_{k} } + \beta_{k} e^{\varpi(1-\frac{\beta_{k}}{K-1}) }], \\
\end{aligned}
\end{equation}
where $L(\varpi, X) = \sum_{x_i} p(x_i) e^{\varpi\{1-p(x_i)\}}$.

Then, by using Lagrange multiplier method, we have
\begin{equation}
\begin{aligned}
    & G(p(x_i), \lambda_0)\\
    & = \sum\limits_{x_i} p(x_i)e^{\varpi(1-p(x_i))} + \lambda_0 \big[\sum\limits_{x_i} p(x_i)-1 \big].
\end{aligned}
\end{equation}

By setting $\frac{\partial G(p(x_i),\lambda_0)}{\partial p(x_i)}=0$ and $\frac{\partial G(p(x_i),\lambda_0)}{\partial \lambda_0}=0$, it can be readily verified that the extreme value of $\sum_{y_j} p(y_j)e^{\varpi(1-p(y_j))}$ is achieved by the uniform distribution as a solution, that is $p(x_1)=p(x_2)=...=p(x_K)=1/K$.
In the case that $0<\varpi < 2 \le 2/\max{\{ p(x_i) \} }$, we have $\frac{\partial^2 G(p(x_i),\lambda_0)}{\partial p^2(x_i)}<0$ with respect to $p(x_i) \in [0,1]$, which implies that extreme value of $\sum_{x_i} p(x_i)e^{\varpi(1-p(x_i))}$ is the maximum value.

In addition, according to the Eq. (\ref{eq.p_x}), the uniform distribution of variable $X$ is resulted from the uniform distribution for variable $Y$.

Therefore, by substituting the uniform distribution for $p(x)$ into Eq. (\ref{eq.C_derivative}), we will obtain the capacity $C(\varpi, \beta_{k})$.
\end{IEEEproof}

Furthermore, in light of Eq. (\ref{eq.C_strongly}), we have
\begin{equation}
\begin{aligned}
    & \frac{\partial C(\varpi,\beta_{k})}{\partial \beta_{k}} \\
    & = \{ 1-\varpi(1-\beta_{k})\}e^{\varpi \beta_{k} } + \Big\{ \frac{\varpi\beta_{k}}{K-1}-1\Big\}e^{\varpi(1-\frac{\beta_{k}}{K-1}) },
\end{aligned}
\end{equation}
By setting $\frac{\partial C(\varpi,\beta_{k})}{\partial \beta_{k}}=0$, it is apparent that $C(\varpi,\beta_{k})$ reaches the extreme value in the case that $\beta_{k}= \frac{K-1}{K}$.
Additionally, when the parameter $\varpi$ satisfies $0< \varpi < 2 \le 2/\max{\{ p(x_i)\}}$, we also have the second derivative of the $C(\varpi,\beta_{k})$ as follows
\begin{equation}
\begin{aligned}
    & \frac{\partial^2 C(\varpi,\beta_{k})}{\partial \beta_{k}^2} \\
    & =  \varpi[2 - (1-\beta_{k})\varpi ]e^{\varpi \beta_{k} } +
     \frac{\varpi}{K-1}\Big\{ 2 - \frac{\varpi\beta_{k}}{K-1} \Big\}e^{\varpi(1-\frac{\beta_{k}}{K-1}) }\\
    & > 0,
\end{aligned}
\end{equation}
which indicates the convex $C(\varpi,\beta_{k})$ reaches the minimum value 0 in the case $\beta_{k}= \frac{K-1}{K}$.

\begin{rem}
According to Proposition \ref{prop_3}, when $\beta_{k}=\frac{K-1}{K}$, namely, the channel is just random, we gain the lower bound of the capacity namely $C(\varpi, \beta_{k})=0$. On the contrary, when $\beta_{k}=0$, that is, there is a certain channel, we will have the maximum capacity.
\end{rem}

\section{Distortion of message importance transfer }
In this section, we will focus on the information transfer distortion, a common problem of information processing.
In a real information system, there exists inevitable information distortion caused by noises or other disturbances,
though the devices and hardware of telecommunication systems are updating and developing.
Fortunately, there are still some bonuses from allowable distortion in some scenarios. For example, in conventional information theory, rate distortion is exploited to obtain source compression such as predictive coding and
hybrid encoding, which can save a lot of hardware resources and communication traffic \cite{On-the-rate-distortion-function}.

Similar to the rate distortion theory for Shannon entropy \cite{Elements-of-information-theory-2nd-edition}, a kind of rate distortion function based on MIM and CMIM is defined to characterize the effect of distortion on the message importance loss. In particular, there are some details of discussion as follows.

\begin{defn}\label{defn.RD}
Assume that there exists an information transfer process $ \{ X, p(y|x), Y\}$ from the variable $X$ to $Y$,
where the $p(y|x)$ denotes a transfer matrix (distributions of $X$ and $Y$ are denoted by $p(x)$ and $p(y)$ respectively).
For a given distortion function $d(x,y)$ ($d(x,y)\ge 0$) and an allowable distortion $D$, the message importance distortion function is defined as
\begin{equation}\label{eq.RD}
\begin{aligned}
     R_{\varpi}(D)
    & = \min_{p(y|x)\in B_D} \Phi_{\varpi}(X||Y)\\
    & = \min_{p(y|x)\in B_D} \{L(\varpi,X)-L(\varpi,X|Y)\},\\
\end{aligned}
\end{equation}
in which
$L(\varpi, X) = \sum_{x_i} p(x_i) e^{\varpi\{1-p(x_i)\}}$, $L(\varpi, X|Y)$ is defined by Eq. (\ref{CMIM_discrete1}), $0<\varpi \le \frac{2\min_j{\{p(y_j)\}}}{ \max_i{\{ p(x_i) \}} }$ and $B_D=\{q(y|x):\bar D \le D\}$ denotes the allowable information transfer matrix set where
\begin{equation}\label{eq.ave_distortion}
\begin{aligned}
    & \bar D = \sum_{x_i}\sum_{y_j}p(x_i)p(y_j|x_i) d(x_i,y_j),
\end{aligned}
\end{equation}
which is the average distortion.
\end{defn}

In this model, the information source $X$ is given  and our goal is to select an adaptive $p(y|x)$ to means the minimum allowable message importance loss under the distortion constraint. This provides a new theoretical guidance for information source compression from the perspective of rare events semantics.

In contrast to the rate distortion of Shannon information theory, this new information distortion function just depends on the message importance loss rather than entropy loss to choose an appropriate information compression matrix.
In practice, there are some similarities and differences between these two rate distortion theories for source compression.
On one hand, both two rate distortion encodings can be regarded as special information transfer processes just with different optimization objectives.
On the other hand, the new distortion theory tries to keep the rare probability events as many as possible, while the conventional rate distortion focuses on the amount of information itself. To some degree, by reducing more redundant common information, the new source compression strategy based on rare events (viewed as message importance) may save more computing and storage resources in big data.

\subsection{ Properties of message importance distortion function}\label{section_RD1}
In this subsection, we shall discuss some fundamental properties of rate distortion function based on message importance in details.
\subsubsection{Domain of distortion}\ \par
Here we investigate the domain of allowable distortion, namely $[D_{\min}, D_{\max}]$, and the corresponding message importance distortion function values as follows.

\textit{i)} The lower bound $D_{\min}$: Due to the fact $0 \le d(x_i,y_j)$, it is easy to obtain the non-negative average distortion, namely $0 \le \bar D $. Considering $\bar D\le D$, we readily have the minimum allowable distortion, that is
\begin{equation}
 \begin{aligned}
    D_{\min} = 0,
 \end{aligned}
\end{equation}
which implies the distortionless case, namely $Y$ is the same as $X$.

In addition, when the lower bound $D_{\min}$ (namely the distortionless case) is satisfied, it is readily to see that
\begin{equation}
 \begin{aligned}
     L(\varpi, X|Y)
     & = L(\varpi, X|X) \\
     & = \sum_{x_i} p(x_i) p(x_i|x_i)e^{\varpi\{1-p(x_i|x_i)\}}
     = 1,
 \end{aligned}
\end{equation}
and according to the Eq. (\ref{eq.RD}) the message importance distortion function is
\begin{equation}
 \begin{aligned}
     R_{\varpi}(D_{\min}) & = L(\varpi, X) - L(\varpi, X|X)\\
     & = L(\varpi, X) - 1,
 \end{aligned}
\end{equation}
where $L(\varpi, X) = \sum_{x_i} p(x_i) e^{\varpi\{1-p(x_i)\}}$ and $0<\varpi \le \frac{2\min_j{\{p(y_j)\}}}{ \max_i{\{ p(x_i) \}} }$.

\textit{ii)} The upper bound $D_{\max}$: When the allowable distortion satisfies $D \ge D_{\max}$, it is apparent that the variables $X$ and $Y$ are independent, that is, $p(y|x) = p(y)$. Furthermore, it is readily to see that
\begin{equation}
 \begin{aligned}
     D_{\max} & = \min_{p(y)} \big\{ \sum_{x_i}\sum_{y_j} p(x_i)p(y_j)d(x_i,y_j)\big\}\\
     & = \sum_{y_j}p(y_j) \min_{p(y)} \big\{ \sum_{x_i} p(x_i)d(x_i,y_j)\big\}\\
     & \ge \min_{y_j} \big\{ \sum_{x_i} p(x_i)d(x_i,y_j)\big\}
 \end{aligned}
\end{equation}
which indicates that when the distribution of variable $Y$ follows $p(y_j)=1$ and $p(y_l)=0$ ($l\ne j$), we have the upper bound
\begin{equation}\label{eq.domainDmax}
 \begin{aligned}
     D_{\max}=\min_{y_j} \big\{ \sum_{x_i} p(x_i)d(x_i,y_j)\big\}.
 \end{aligned}
\end{equation}

Additionally, on account of the independent $X$ and $Y$, namely $p(x|y)=p(x)$, it is readily to see that
\begin{equation}\label{eq.RDmax}
 \begin{aligned}
     R_{\varpi}(D_{\max}) & = L(\varpi, X) - \sum_{y_j}p(y_j)L(\varpi, X)
      = 0.
 \end{aligned}
\end{equation}

\subsubsection{The convexity property}\ \par
For two allowable distortions $D_a$ and $D_b$, whose optimal allowable information transfer matrix are $p_a(y|x)$ and $p_b(y|x)$ respectively, we have
\begin{equation}\label{eq.convexity}
 \begin{aligned}
    & R_{\varpi}(\delta D_a + (1-\delta) D_b) \\
    & \qquad \qquad \quad \le \delta R_{\varpi}(D_a) + (1-\delta)R_{\varpi}(D_b),
 \end{aligned}
\end{equation}
where $0\le \delta \le 1$ and $0<\varpi \le \frac{2\min_j{\{p(y_j)\}}}{ \max_i{\{ p(x_i) \}} }$.

\begin{IEEEproof}
As for an allowable distortion $D_0 =\delta D_a +(1-\delta)D_b$, we have the average distortion for the information transfer matrix $p_0(y|x)=\delta p_a(y|x) +(1-\delta) p_b(y|x)$ as follows
\begin{equation}
 \begin{aligned}
     \bar D_0
    &=  \delta\sum_{x_i}\sum_{y_j}p(x_i)p_a(y_j|x_i) d(x_i,y_j) \\
    & \quad + (1-\delta)\sum_{x_i}\sum_{y_j}p(x_i)p_b(y_j|x_i) d(x_i,y_j)\\
    & \le \delta {D_a} +(1-\delta) {D_b} = D_0,
 \end{aligned}
\end{equation}
which indicates the $p_0(y|x)$ is an allowable information transfer matrix for $D_0$.

Moreover, by using Jensen's inequality and Bayes' theorem, we have the CMIM with respect to $p_0(y|x)$ as the Eq. (\ref{eq.L0})
\begin{figure*}
\begin{equation}\label{eq.L0}
 \begin{aligned}
      L_{0}(\varpi, X|Y)
     & = \sum\limits_{x_i}\sum\limits_{y_i} p(x_i)p_0(y_j|x_i) e^{\varpi\{1-\frac{p(x_i)p_0(y_j|x_i)}{p_0(y_j)}\}}\\
     & =  \sum\limits_{x_i}\sum\limits_{y_i} p(x_i) [\delta p_a(y_j|x_i)+(1-\delta)p_b(y_j|x_i)]
     e^{\varpi\{1-\frac{p(x_i)[\delta p_a(y_j|x_i)+(1-\delta)p_b(y_j|x_i)]}{p_0(y_j)}\}}\\
     & \ge \sum\limits_{x_i}\sum\limits_{y_i} p(x_i) [\delta p_a(y_j|x_i)]
     e^{\varpi\{1-\frac{p(x_i)[\delta p_a(y_j|x_i)]}{p_0(y_j)}\}}
      + \sum\limits_{x_i}\sum\limits_{y_i} p(x_i) [(1-\delta)p_b(y_j|x_i)]
     e^{\varpi\{1-\frac{p(x_i)[(1-\delta)p_b(y_j|x_i)]}{p_0(y_j)}\}}\\
     & \ge \delta\sum\limits_{x_i}\sum\limits_{y_i} p(x_i) p_a(y_j|x_i)
     e^{\varpi\{1-\frac{p(x_i)p_a(y_j|x_i)}{p_a(y_j)}\}}
      + (1-\delta)\sum\limits_{x_i}\sum\limits_{y_i} p(x_i)p_b(y_j|x_i)
     e^{\varpi\{1-\frac{p(x_i)p_b(y_j|x_i)}{p_b(y_j)}\}}\\
     & = \delta L_{a}(\varpi, X|Y) + (1-\delta) L_{b}(\varpi, X|Y),
 \end{aligned}
\end{equation}
\hrulefill
\vspace*{4pt}
\end{figure*}
in which
\begin{equation}
 \begin{aligned}
    p_0(y_j)
    & = \sum_{x_i}p(x_i)p_0(y_j|x_i)\\
    & = \sum_{x_i}p(x_i)[\delta p_a(y_j|x_i)+(1-\delta)p_b(y_j|x_i)]\\
    & = \delta p_a(y_j)+(1-\delta)p_b(y_j),
 \end{aligned}
\end{equation}
and the parameter $\varpi$ is $0<\varpi \le \frac{2\min_j{\{p(y_j)\}}}{ \max_i{\{ p(x_i) \}} }$.

Furthermore, according to the Eq. (\ref{eq.RD}) and Eq. (\ref{eq.L0}), it is not difficult to have
\begin{equation}
 \begin{aligned}
     R_{\varpi}(D_0)
    & = \min_{p(y|x)\in B_{D_0}} \{L(\varpi,X)-L(\varpi,X|Y)\}\\
    & \le \{L(\varpi,X)-L_0(\varpi,X|Y)\}\\
    & \le \delta\{ L(\varpi,X)-L_a(\varpi,X|Y)\}\\
    & \quad +(1-\delta)\{ L(\varpi,X)-L_b(\varpi,X|Y)\}\\
    & = \delta R_{\varpi}(D_a)+R_{\varpi}(D_b),
 \end{aligned}
\end{equation}
where $L(\varpi,X)$ is the MIM for the given information source $X$, while $L_a(\varpi,X|Y)$ and $L_b(\varpi,X|Y)$ denote the CMIM with respect to $p_a(y|x)$ and $p_b(y|x)$ respectively.

Therefore, the convexity property is tesitfied.
\end{IEEEproof}

\subsubsection{The monotonically decreasing property}\ \par
For two given allowable distortions $D_a$ and $D_b$, if $0\le D_a < D_b<D_{\max}$ is satisfied, we have
$R_{\varpi}(D_a) \ge R_{\varpi}(D_b)$,
where $0<\varpi \le \frac{2\min_j{\{p(y_j)\}}}{ \max_i{\{ p(x_i) \}} }$.
\begin{IEEEproof}
Considering that $0\le D_a<D_b< D_{\max}$, we have
$D_b= \gamma D_a + (1-\gamma)D_{\max}$ where $\gamma= \frac{D_{\max}-D_b}{D_{\max}-D_a}$. On account of the Eq. (\ref{eq.RDmax}) and the convexity property mentioned in the Eq. (\ref{eq.convexity}), it is not difficult to see that
\begin{equation}
 \begin{aligned}
     R_{\varpi}(D_b)
     & \le \gamma R_{\varpi}(D_a) +(1-\gamma)R_{\varpi}(D_{\max}) \\
     & \qquad \qquad \qquad \quad = \gamma R_{\varpi}(D_a) < R_{\varpi}(D_a),
 \end{aligned}
\end{equation}
which verifies this property.
\end{IEEEproof}

\subsubsection{The equivalent expression}\ \par
For an information transfer process $ \{ X, p(y|x), Y\}$, if we have a given distortion function $d(x,y)$, an allowable distortion $D$ and a average distortion $\bar D$ defined in the Eq. (\ref{eq.ave_distortion}), the message importance distortion function defined in the Eq. (\ref{eq.RD}) can be rewritten as
\begin{equation}\label{eq.RD_equal}
\begin{aligned}
    & R_{\varpi}(D) = \min_{\bar D = D} \{L(\varpi,X)-L(\varpi,X|Y)\},\\
\end{aligned}
\end{equation}
where $L(\varpi, X)$ and $L(\varpi, X|Y)$ are defined by the Eq. (\ref{defn:MIM}) and Eq. (\ref{CMIM_discrete1}), as well as $0<\varpi \le \frac{2\min_j{\{p(y_j)\}}}{ \max_i{\{ p(x_i) \}} }$.

\begin{IEEEproof}
For a given allowable distortion $D$, if there exists an allowable distortion $D^*$ ($D_{\min}\le D^{*}<D < D_{\max} $) and the corresponding optimal information transfer matrix $p^*(y|x)$ leads to $R_{\varpi}(D)$, we will have
$R_{\varpi}(D) = R_{\varpi}(D*)$ which contradicts the monotonically decreasing property.
Thus, the proof of this property is completed.
\end{IEEEproof}

\subsection{ Analysis for message importance distortion function}
In this subsection, we shall investigate the computation of message importance distortion function, which has a great impact on the probability events analysis in practice.
Actually, the definition of message importance distortion function in the Eq. (\ref{eq.RD}) can be regarded as a special function, which is the minimization of the message importance loss with the symbol error less than or equal to the allowable distortion $D$.
In particular, the Definition \ref{defn.RD} can also be expressed as the following optimization
\begin{flalign}\label{eq.RD optization}
\mathcal{P}_1: \mathop{\min}\limits_{p(y_j|x_i)}\,\,\, &\{ L(\varpi, X) - L(\varpi, X|Y)\}  \\
\textrm{s.t.}\,\,\,
& \sum_{x_i}\sum_{y_j} p(x_i)p(y_j|x_i)d(x_i,y_j)\le D, \tag{\theequation a}\label{equ:LC1_optization_a}\\
& \sum_{y_j} p(y_j|x_i)=1, \tag{\theequation b}\label{equ:LC1_optization_b}\\
& p(y_j|x_i) \ge 0 ,\tag{\theequation c}\label{equ:LC1_optization_c}
\end{flalign}
where $L(\varpi, X)$ and $L(\varpi, X|Y)$ are MIM and CMIM defined in the Eq. (\ref{defn:MIM}) and Eq. (\ref{CMIM_discrete1}), as well as $0<\varpi \le \frac{2\min_j{\{p(y_j)\}}}{ \max_i{\{ p(x_i) \}} }$.

To take a computable optimization problem as an example, we consider Hamming distortion as the distortion function $d(x,y)$, namely
\begin{equation}\label{eq.dxy_hamming}
\begin{aligned}
    d(x,y) = \left [
    \begin{matrix}
        0 & 1 & ... & 1\\
        1 & 0 & ... & 1 \\
        ... & ... &  & ...\\
        1 & 1 & ... & 0
    \end{matrix}
    \right ],
\end{aligned}
\end{equation}
which means $d(x_i,y_i)=0$ and $d(x_i,y_j)=1$ ($i\ne j$).
In order to reveal some intrinsic meanings of $R_{\varpi}(D)$, we investigate an information transfer of Bernoulli source as follows.

\begin{prop}\label{prop.binary_RD}
For a Bernoulli($p$) source denoted by a variable $X$ and an information transfer process $\{X, p(y|x), Y\}$ with Hamming distortion, the message importance distortion function is given by
\begin{flalign}\label{equ:prop_Rate}
\begin{aligned}
&{R_{\varpi}}(D)
= \{p e^{\varpi(1-p)} + (1-p)e^{\varpi p}\} \\
& \qquad \qquad \qquad \qquad - \{D e^{\varpi(1-D)} + (1-D)e^{\varpi D}\},
\end{aligned}
\end{flalign}
and the corresponding information transfer matrix is
\begin{equation}\label{eq.transfer_matrixRD}
\begin{aligned}
    p(y|x) = \left [
    \begin{matrix}
        \frac{(1-D)(p-D)}{p(1-2D)} & \frac{(1-p-D)D}{p(1-2D)} \\
        \frac{D(p-D)}{(1-p)(1-2D)} & \frac{(1-p-D)(1-D)}{(1-p)(1-2D)} \\
    \end{matrix}
    \right ],
\end{aligned}
\end{equation}
where $0<\varpi \le \frac{2\min_j{\{p(y_j)\}}}{ \max_i{\{ p(x_i) \}} }$ and $0\le D \le \min\{p,1-p\}$.
\end{prop}
\begin{IEEEproof}
Considering the fact that the Bernoulli source $X$ is given and the equivalent expression is mentioned in the Eq. (\ref{eq.RD_equal}),
the optimization problem $\mathcal{P}_1$ can be regarded as
\begin{flalign}\label{eq.RD_A optization}
\mathcal{P}_{1-A}:\mathop{\max}\limits_{p(y_j|x_i)} &\,\,\, L(\varpi, X|Y)  \\
\textrm{s.t.}\,\,\,
& p(x_0)p(y_1|x_0)+p(x_1)p(y_0|x_1)= D, \tag{\theequation a}\label{equ:RD_A_optization_a}\\
& p(y_0|x_0)+p(y_1|x_0) =1,  \tag{\theequation b}\label{equ:RD_A_optization_b}\\
& p(y_0|x_1)+p(y_1|x_1) =1, \tag{\theequation c}\label{equ:RD_A_optization_c}\\
& p(y_j|x_i) \ge 0, \quad (i=0,1;j=0,1), \tag{\theequation d}\label{equ:RD_A_optization_d}
\end{flalign}
where $L(\varpi, X|Y)=\sum_{x_i,y_j} p(x_i,y_j)e^{\varpi(1-p(x_i|y_j))}$ and $0<\varpi \le \frac{2\min_j{\{p(y_j)\}}}{ \max_i{\{ p(x_i) \}} }$.

To simplify the above one, we have
\begin{flalign}\label{eq.RD_B optimization}
\mathcal{P}_{1-B}:\mathop{\max}\limits_{\alpha, \beta} &\,\,\, L_{D}(\varpi, X|Y)  \\
\textrm{s.t.}\,\,\,
& p\alpha +(1-p)\beta= D, \tag{\theequation a}\label{equ:RD_B_optization_a}\\
& 0 \le \alpha \le 1, 0 \le \beta \le 1,0\le p\le 1, \tag{\theequation b}\label{equ:RD_B_optization_b}
\end{flalign}
in which $p$ and $(1-p)$ denote $p(x_0)$ and $p(x_1)$, $\alpha$ and $\beta$ denote $p(y_1|x_0)$ and $p(y_0|x_1)$, and
\begin{equation}
\begin{aligned}
& L_{D}(\varpi, X|Y)\\
& = p(1-\alpha)e^{\frac{\varpi(1-p)\beta}{p(1-\alpha)+(1-p)\beta}}+(1-p)\beta e^{\frac{\varpi p(1-\alpha)}{(1-p)\beta+ p(1-\alpha)}}\\
& \quad + (1-p)(1-\beta)e^{\frac{\varpi p\alpha}{p\alpha+(1-p)(1-\beta)}}+p\alpha e^{\frac{\varpi(1-p)(1-\beta)}{p\alpha+(1-p)(1-\beta)}},
\end{aligned}
\end{equation}
where $0<\varpi \le \frac{2\min_j{\{p(y_j)\}}}{ \max_i{\{ p(x_i) \}} }$.

Actually, it is not easy to deal with the Eq. (\ref{eq.RD_B optimization}) directly, we intend to use an equivalent expression to describe this objective. By using Taylor series expansion of $e^x$, namely $e^x = 1+ x + \frac{x^2}{2}+o(x^2)$, we have
\begin{equation}\label{eq.LDxy}
\begin{aligned}
& L_{D}(\varpi, X|Y)\\
& \doteq 1+ (2\varpi + \frac{\varpi^2}{2}) \Big\{\frac{p\alpha (1-p)(1-\beta) }{p\alpha + (1-p)(1-\beta)} \\
& \qquad \qquad \qquad \qquad \qquad + \frac{p(1-\alpha) (1-p)\beta }{p(1-\alpha)+(1-p)\beta } \Big\}.
\end{aligned}
\end{equation}
By substituting $\beta = \frac{D-p\alpha}{1-p}$ into the Eq. (\ref{eq.LDxy}), it is easy to have
\begin{equation}\label{eq.LDxy_2}
\begin{aligned}
& L_{D}(\varpi, X|Y)\\
& \doteq 1+ p(2\varpi + \frac{\varpi^2}{2})  \Big\{\frac{p\alpha^2 + (1-p-D)\alpha }{2p\alpha+(1-p-D)} \\
& \qquad \qquad \qquad \qquad \qquad + \frac{p\alpha^2 - (p+D)\alpha +D}{(p+D) - 2p\alpha} \Big\},
\end{aligned}
\end{equation}
where $\max\{0,1+\frac{D-1}{p}\} \le \alpha \le \min\{1,\frac{D}{p}\}$ resulted from the constraints in the Eq. (\ref{equ:RD_B_optization_a}) and Eq. (\ref{equ:RD_B_optization_b}).

Moreover, it is not difficult to have the partial derivative of $L_{D}(\varpi, X|Y)$ in the Eq. (\ref{eq.LDxy_2}) with respect to $\alpha$ as follows
\begin{equation}\label{eq.dLDxy_2}
\begin{aligned}
& \frac{\partial L_{D}(\varpi, X|Y)}{\partial \alpha}\\
& \doteq 2p^2(2\varpi + \frac{\varpi^2}{2})  \Big\{\frac{-p\alpha^2 - (1-p-D)\alpha }{[2p\alpha+(1-p-D)]^2} \\
& \qquad \qquad \qquad \qquad \qquad + \frac{p\alpha^2 - (p+D)\alpha +D}{[(p+D) - 2p\alpha]^2} \Big\}.
\end{aligned}
\end{equation}
By setting $\frac{\partial L_{D}(\varpi, X|Y)}{\partial \alpha}=0$, it is not difficult to see that the solutions of $\alpha$ in the Eq. (\ref{eq.dLDxy_2}) are given by
$\alpha_1 = \frac{(1-p-D)D}{p(1-2D)}$ and $\alpha_2 = \frac{1-D-p}{1-2p}$ respectively.

In addition, in the light of the domain of $D$ mentioned in the Eq. (\ref{eq.domainDmax}), it is readily to have $D_{\max} = \min\{p,1-p\}$ in the Bernoulli source case. That is, the allowable distortion satisfies $0\le D\le \min\{p, 1-p\}$. Thus, the domain of $\alpha$ namely $\max\{0,1+\frac{D-1}{p}\} \le \alpha \le \min\{1,\frac{D}{p}\}$, can be given by $0 \le \alpha \le \frac{D}{p}$.

Then, it is readily to have the appropriate solution of $\alpha$ as follows
\begin{equation}\label{eq.alpha}
\begin{aligned}
\alpha^* = \frac{(1-p-D)D}{p(1-2D)},
\end{aligned}
\end{equation}
in which the second derivative $\frac{\partial^2 L_{D}(\varpi, X|Y)}{\partial \alpha^2}$ is not positive, namely maximum value is reached, and the corresponding information transfer matrix is
\begin{equation}\label{eq.transfer_matrix}
\begin{aligned}
    p(y|x) = \left [
    \begin{matrix}
        \frac{(1-D)(p-D)}{p(1-2D)} & \frac{(1-p-D)D}{p(1-2D)} \\
        & \\
        \frac{D(p-D)}{(1-p)(1-2D)} & \frac{(1-p-D)(1-D)}{(1-p)(1-2D)} \\
    \end{matrix}
    \right ],
\end{aligned}
\end{equation}
where $0\le D\le \min\{p, 1-p\}$.

Consequently, by substituting the matrix Eq. (\ref{eq.transfer_matrix}) into the Eq. (\ref{eq.RD optization}), it is not difficult to verify this proposition.
\end{IEEEproof}


\section{ Bitrate transmission constrained by message importance}
We investigate information capacity in the case of a limited message importance loss in this section. The objective is to achieve the maximum transmission bitrate under the constraint of a certain message importance loss $\epsilon$. The maximum transmission bitrate is one of system invariants in a transmission process, which provides a upper bound of amount of information obtained by the receiver.

In an information transmission process satisfying the AEP condition, the information capacity is the mutual information between the encoded signal and the received signal with the dimension bit/symbol.
In a real transmission, there always exists an allowable distortion between the sending sequence $X$ and the received sequence $Y$, while the maximum allowable message importance loss is required to avoid too much distortion of rare events.
From this perspective, message importance loss is considered to be another constraint for the information transmission capacity beyond the information distortion.
Therefore, this might play a crucial role in the design of transmission in information processing systems.

In particular, we characterize the maximizing mutual information constrained by a controlled message importance loss as follows
\begin{flalign}\label{equ:IL2 optization}
\mathcal{P}_2: \,\,\mathop {\max }\limits_{p(x)}\,\,\, &I(X||Y)  \\
\textrm{s.t.}\,\,\,
& L(\varpi, X) - L(\varpi, X|Y)  \le \epsilon \tag{\theequation a}\label{equ:IL1 optization a},\\
& \sum_{y_j} p(x_i)=1, \tag{\theequation b}\label{equ:IL1 optization b}\\
& p(x_i) \ge 0 ,\tag{\theequation c}\label{equ:IL1 optization c}
\end{flalign}
where $I(X||Y)= \sum_{x_i,y_j}p(x_i)p(y_j|x_i) \log \frac{p(x_i)p(y_j|x_i)}{p(y_j)}$, $p(y_j)= \sum_{x_i} p(x_i)p(y_j|x_i)$, $L(\varpi, X)$ and $L(\varpi, X|Y)$ are MIM and CMIM defined in the Eq. (\ref{defn:MIM}) and Eq. (\ref{CMIM_discrete1}), as well as $0<\varpi \le \frac{2\min_j{\{p(y_j)\}}}{ \max_i{\{ p(x_i) \}} }$.

Actually, the bitrate transmission with a message importance loss constraint has a special solution for a certain scenario.
In order to give a specific example, we investigate the optimization problem in the Bernoulli($p$) source with the symmetric or erasure transfer matrix as follows.

\subsection{Binary symmetric matrix}
\begin{prop}\label{prop.symmetric_IL}
For a Bernoulli(p) source $X$ whose distribution is $\{p,1-p\}$ ($0\le p\le 1/2$) and an information transfer process $\{ X, p(y|x), Y\}$ with transfer matrix
\begin{equation}\label{eq.symmetric_channel}
 \begin{aligned}
   p(y|x) = \left [
   \begin{matrix}
    1-\beta_{s} & \beta_{s} \\
    \beta_{s} & 1-\beta_{s}
   \end{matrix}
   \right ],
 \end{aligned}
\end{equation}
we have the solution for $\mathcal{P}_2$ defined in the Eq. (\ref{equ:IL2 optization}) as follows
\begin{equation}
\begin{split}
 & \mathop {\max }\limits_{p(x)} I(X||Y)\\
 &=
  \left\{
   \begin{aligned}
   &1- H(\beta_s), \,\, ( \epsilon \ge C_{\beta_s})\\
   &H(p_s(1-\beta_s)+(1-p_s)\beta_s)- H(\beta_s),
   \,\, ( 0< \epsilon \le C_{\beta_s})
   \end{aligned}
   \right.
   \end{split}
\end{equation}
where $p_s$ is the solution of $L(\varpi, X) - L(\varpi, X|Y)  = \epsilon$ ($L(\varpi, X)$ and $L(\varpi, X|Y)$ mentioned in the optimization problem $\mathcal{P}_2$), whose approximate value is
\begin{equation}\label{eq.p_s}
 \begin{aligned}
    p_s \doteq \frac{1-\sqrt{\Theta}}{2},
 \end{aligned}
\end{equation}
in which the parameter $\Theta$ is given by
\begin{equation}
 \begin{aligned}
    \Theta
    & = 1- \frac{4\epsilon}{4\varpi+{\varpi^2}}\\
    & \quad- \frac{4\sqrt{ (1-2\beta_s)^2\epsilon^2 + 2(4\varpi+\varpi^2)\beta_s(1-\beta_s)\epsilon}}{(4\varpi+{\varpi^2})|1-2\beta_s|},
 \end{aligned}
\end{equation}
and $H(\cdot)$ denotes the operator for Shannon entropy, that is $H(p)=-[(1-p)\log(1-p)+p\log p]$ , $C_{\beta_s} = e^{ \frac{\varpi}{2} } - \{\beta_{s}  e^{\varpi(1-\beta_{s}) } + (1-\beta_{s} )e^{\varpi \beta_{s}}\} $ ($0\le \beta_{s} \le 1$) and $\varpi < 2 \le 2/\max{\{ p(x_i) \} }$.
\end{prop}

\begin{IEEEproof}
Considering the Bernoulli($p$) source $X$ following $\{p,1-p\}$ and the binary symmetric matrix, it is not difficult to gain
\begin{equation}\label{eq.Ixy_binary}
 \begin{aligned}
  I(X||Y)
 & = H(Y)- H(Y|X)\\
 & = -\{p(y_0)\log p(y_0) + p(y_1)\log p(y_1)\} - H(\beta_s),
 \end{aligned}
\end{equation}
where $p(y_0)= p(1-\beta_s)+(1-p)\beta_s$, $p(y_1)= p\beta_s+(1-p)(1-\beta_s)$ and $H(\beta_s) = -[(1-\beta_s)\log(1-\beta_s)+\beta_s\log \beta_s]$.

Moreover, define the Lagrange function as $G_s(p) = I(X||Y) + \lambda_s (L(\varpi, X)-L(\varpi,X|Y)-\epsilon)$ where $\epsilon >0 $, $0\le p\le 1/2$ and $\lambda_s\ge 0$. It is not difficult to have the partial derivative of $G_s(p)$ as follows
\begin{equation}\label{eq.Ixy}
 \begin{aligned}
  & \frac{\partial G_s(p)}{\partial p}
  =  \frac{\partial I(X||Y)}{\partial p} + \lambda_s\frac{\partial C(p,\varpi, \beta_{s} )}{\partial p},
 \end{aligned}
\end{equation}
where $\frac{\partial C(p,\varpi, \beta_{s} )}{\partial p}$ is given by the Eq. (\ref{eq.dCbinary}) and
\begin{equation}\label{eq.dIxy}
 \begin{aligned}
\frac{\partial I(X||Y)}{\partial p} = (1-2\beta_s) \log\bigg\{\frac{(2\beta_s-1)p+1-\beta_s}{(1-2\beta_s)p+\beta_s}\bigg\}.
 \end{aligned}
\end{equation}

By virtue of the monotonic increasing function $\log(x)$ for $x>0$, it is easy to see the nonnegativity of $\frac{\partial I(X||Y)}{\partial p} $ is equal to $(1-2\beta_s) \{{(2\beta_s-1)p+1-\beta_s}-[{(1-2\beta_s)p+\beta_s}]\}= (1-2p)(1-2\beta_s)^2\ge 0 $ in the case $0\le p\le 1/2$.
Moreover, due to the nonnegative $\frac{\partial C(p,\varpi, \beta_{s} )}{\partial p}$ in $p\in [0,1/2]$ which is mentioned in the proof of Proposition \ref{prop.symmetric}, it is readily seen that $\frac{\partial G_s(p)}{\partial p}\ge 0 $ is satisfied under the condition $0\le p\le 1/2$.

Thus, the optimal solution $p_s^*$ is the maximal available $p$ ($p\in [0,1/2]$) as follows
\begin{equation}
\begin{split}
 \mathop {p_s^*}=
  \left\{
   \begin{aligned}
   &\frac{1}{2}, \,\,\text{for} \quad \epsilon \ge C_{\beta_s},\\
   &p_s, \,\,\text{for} \quad 0< \epsilon \le C_{\beta_s},
   \end{aligned}
   \right.
   \end{split}
\end{equation}
where $p_s$ is the solution of $L(\varpi, X) - L(\varpi, X|Y)  = \epsilon$, and $C_{\beta_s}$ is the MILC mentioned in the Eq. (\ref{eq.C_binary}).

By using Taylor series expansion, the equation $L(\varpi, X) - L(\varpi, X|Y)  = \epsilon$ can be expressed approximately as follows
\begin{equation}
 \begin{aligned}
    & (2\varpi+\frac{\varpi^2}{2})\bigg\{ (1-p)p \\
    & \qquad -\frac{p(1-p)\beta_s(1-\beta_s)}{[{(2\beta_s-1)p+1-\beta_s}][{(1-2\beta_s)p+\beta_s}]} \bigg\}= \epsilon,
 \end{aligned}
\end{equation}
whose solution is the approximate $p_s$ as the Eq. (\ref{eq.p_s}).

Therefore, by substituting the $p_s^*$ into the Eq. (\ref{eq.Ixy_binary}), it is readily to testify the proposition.
\end{IEEEproof}
\begin{rem}\label{rem:symmetric_IL}
Proposition \ref{prop.symmetric_IL} gives the maximum transmission bitrate under the constraint of message importance loss. Particularly, there are growth region and smooth region for the maximum transmission bitrate in the receiver with respect to message importance loss $\epsilon$.
When the message importance loss $\epsilon$ is constrained in a little range, the real bitrate is less than the Shannon information capacity which is concerned with the entropy of the symmetric matrix parameter $\beta_s$.
\end{rem}

\subsection{Binary erasure matrix}
\begin{prop}\label{prop.error_symmetric_IL}
Assume that there is a Bernoulli(p) source $X$ following distribution $\{p,1-p\}$ ($0\le p\le 1/2$) and an information transfer process $\{ X, p(y|x), Y\}$ with the binary erasure matrix
\begin{equation}
 \begin{aligned}
   p(y|x) = \left [
   \begin{matrix}
    1-\beta_{e} & 0 &\beta_{e} \\
    0 & 1-\beta_{e} &\beta_{e}
   \end{matrix}
   \right ],
 \end{aligned}
\end{equation}
where $0\le \beta_e \le 1$. In this case,
the solution for $\mathcal{P}_2$ described in the Eq. (\ref{equ:IL2 optization}) is
\begin{equation}
\begin{split}
 & \mathop {\max }\limits_{p(x)} I(X||Y)\\
 &=
  \left\{
   \begin{aligned}
   &1-\beta_e, \,\, ( \epsilon \ge C_{\beta_e})\\
   &(1-\beta_e)H(p_e),
   \,\, ( 0< \epsilon \le C_{\beta_s})
   \end{aligned}
   \right.
   \end{split}
\end{equation}
where $p_e$ is the solution of $ (1-\beta_{e})\{pe^{\varpi(1-p)}+(1-p)e^{\varpi p} -1\} = \epsilon $, whose approximate value is
\begin{equation}\label{eq.p_e}
 \begin{aligned}
    p_e \doteq \frac{1-\sqrt{1-\frac{8\epsilon}{(1-\beta_e)(4\varpi+\varpi^2)}}}{2},
 \end{aligned}
\end{equation}
and $H(x)=-[(1-x)\log(1-x)+x\log x]$, $C_{\beta_e} =  (1-\beta_{e}) ( e^{\frac{\varpi}{2}} - 1) $ and $\varpi < 2 \le 2/\max{\{ p(x_i) \} }$.
\end{prop}

\begin{IEEEproof}
In the binary erasure matrix, considering the Bernoulli($p$) source $X$ whose distribution is $\{p,1-p\}$ , it is readily seen that
\begin{equation}\label{eq.Ixy_erasure}
 \begin{aligned}
  I(X||Y)
 & = H(Y)- H(Y|X)\\
 & = (1-\beta_e)H(p),
 \end{aligned}
\end{equation}
where $H(\cdot)$ denotes the Shannon entropy operator, namely $H(p) = -[(1-p)\log(1-p)+p\log p]$.

Moreover, according to the Definition \ref{defn:MIM} and \ref{defn:CMIM}, it is easy to see that
\begin{equation}
L(\varpi, X) - L(\varpi, X|Y) =(1-\beta_{e})\{L(\varpi, p) -1\},
\end{equation}
where $L(\varpi, p)= pe^{\varpi(1-p)}+(1-p)e^{\varpi p}$.

Similar to the proof of the Proposition \ref{prop.symmetric_IL} and considering the monotonically increasing $H(p)$ and $L(\varpi, p)$ in $p\in [0,1/2]$, it is not difficult seen that the optimal solution $p_e^*$ is the maximal available $p$ in the case $0\le p\le \frac{1}{2}$, which is given by
\begin{equation}
\begin{split}
 \mathop {p_e^*}=
  \left\{
   \begin{aligned}
   &\frac{1}{2}, \,\,\text{for} \quad \epsilon \ge C_{\beta_e},\\
   &p_e, \,\,\text{for} \quad 0< \epsilon \le C_{\beta_e},
   \end{aligned}
   \right.
   \end{split}
\end{equation}
where $p_e$ is the solution of $(1-\beta_{e})\{L(\varpi, p) -1\} = \epsilon$, and the upper bound $C_{\beta_e}$ is gained in the Eq. (\ref{eq.Cerasure}).

By resorting to Taylor series expansion, the approximate equation for $(1-\beta_{e})\{L(\varpi, p) -1\} = \epsilon$ is given by
\begin{equation}
 \begin{aligned}
    (1-\beta_e)(2\varpi+\frac{\varpi^2}{2})(1-p)p= \epsilon,
 \end{aligned}
\end{equation}
from which the approximate solution $p_e$ in the Eq. (\ref{eq.p_e}) is obtained.

Therefore, this proposition is readily proved by substituting the $p_e^*$ into the Eq. (\ref{eq.Ixy_erasure}).
\end{IEEEproof}

\begin{rem}\label{rem:error_symmetric_IL}
From Proposition \ref{prop.error_symmetric_IL}, there are two regions for the maximum transmission bitrate with respect to message importance loss. The one depends on the message importance loss threshold $\epsilon$.
The other is just related to the erasure matrix parameter $\beta_e$.
\end{rem}

\section{Numerical Results}
This section shall provide numerical results to validate the theoretical results in this paper.

\subsection{The message importance loss capacity}
\begin{figure}[!t]
\centering
\includegraphics[width=3.6in]{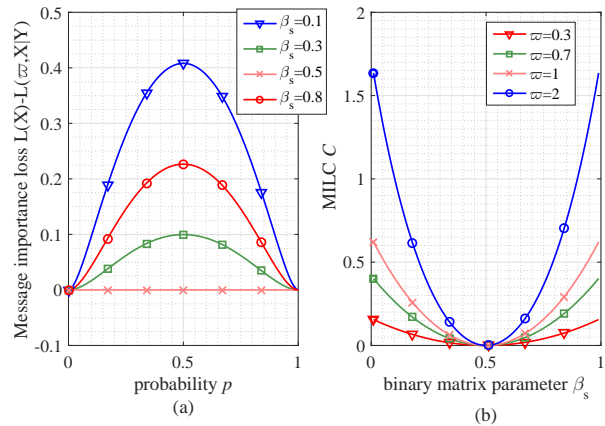}
\caption{The performance of MILC in Binary symmetric matrix. }
\label{fig_2_Capacity}
\end{figure}

First of all, we give some numerical simulation with respect to the MILC in different information transmission cases. In the Fig. \ref{fig_2_Capacity}, it is apparent to see that if the Bernoulli source follows the uniform distribution, namely $p=0.5$, the message importance loss will reach the maximum in the cases of different matrix parameter $\beta_s$. That is, the numerical results of MILC are obtained as $\{0.4081,0.0997,0, 0.2265\}$ in the case of parameter $\beta_s=\{0.1,0.3,0.5,0.8\}$ and $\varpi=1$, which corresponds to the Proposition \ref{prop.symmetric}.
Moreover, we also know that if $\beta_s=0.5$, namely the random transfer matrix is satisfied, the MILC reaches the lower bound that is $C=0$. In the contrast, if the parameter $\beta_s=0$, the upper bound of MILC will be gained such as $\{0.1618,0.4191,0.6487,1.7183\}$ in the case that $\varpi=\{0.3,0.7,1.0,2.0\}$.

\begin{figure}[!t]
\centering
\includegraphics[width=3.6in]{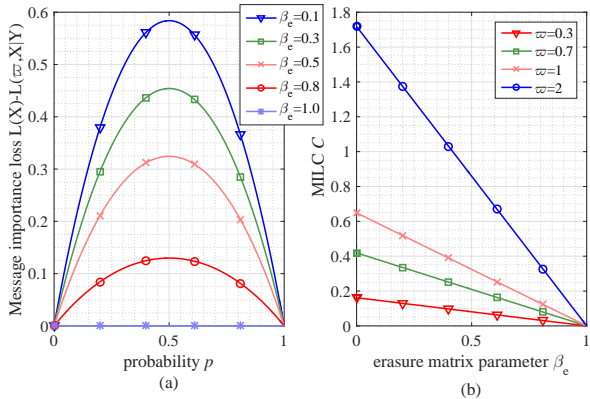}
\caption{The performance of MILC in Binary erasure matrix. }
\label{fig_2error_Capacity}
\end{figure}

Fig. \ref{fig_2error_Capacity} shows that in the transmission with binary erasure matrix, the MILC is reached at the same condition as that with binary symmetric matrix, namely $p=0.5$. For example the numerical results of MILC with $\varpi=1$ are $\{0.5838, 0.4541, 0.3244, 0.1297\}$ in the cases $\beta_e=\{0.1, 0.3, 0.5, 0.8\}$. However, if $\beta_e=1$, the lower bound of MILC ($C=0$) is obtained in the erasure transfer matrix, different from the symmetric case.

\begin{figure}[!t]
\centering
\includegraphics[width=3.4in]{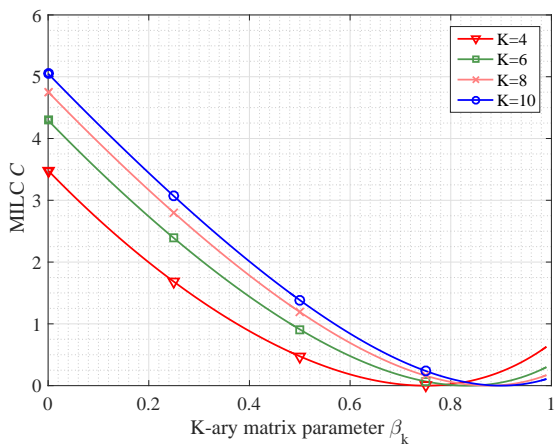}
\caption{The performance of MILC in strongly symmetric matrix with $K=4,6,8,10$. }
\label{fig_K_Capacity}
\end{figure}

From Fig. \ref{fig_K_Capacity}, it is not difficult to see that the certain transfer matrix (namely $\beta_k=0$) leads to upper bound of MILC. For example, when the number of source symbols satisfies $K=\{4,6,8,10\}$, the numerical results of MILC with $\varpi=2$ are $\{ 3.4817,4.2945, 4.7546,5.0496 \}$. Besides, the lower bound of MILC is reached in the case that $\beta_k=1-\frac{1}{K}$.

\subsection{Message importance distortion }
\begin{figure}[!t]
\centering
\includegraphics[width=3.0in]{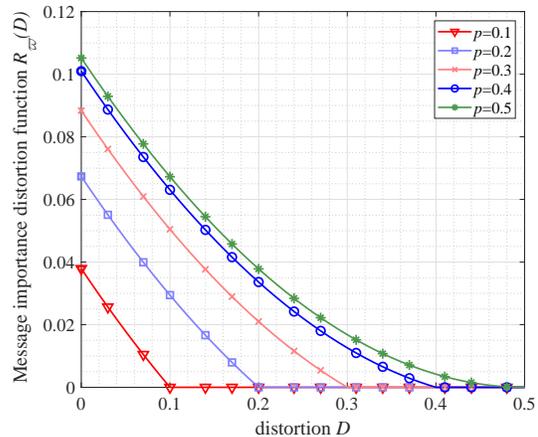}
\caption{The performance of message importance distortion function $R_{\varpi}(D)$ in the case of Bernoulli($p$) source ($p=0.1, 0.2, 0.3, 0.4$). }
\label{fig_RD_Capacity}
\end{figure}

We focus on the distortion of message importance transfer and give some simulations in this subsection.
From Fig. \ref{fig_RD_Capacity}, it is illustrated that the message importance distortion function $R_{\varpi}(D)$ is monotonically non-increasing with respect to the distortion $D$, which can validate some properties mentioned in Section \ref{section_RD1}. Moreover, the maximum $R_{\varpi}(D)$ is obtained in the case $D=0$. Taking the Bernoulli($p$) source as an example, the numerical results of $R_{\varpi}(D)$ with $\varpi=0.2$ are $\{0.0379,0.0674,0.0884,0.1010,0.1052\}$ and the corresponding probability satisfies $p=\{0.1,0.2,0.3,0.4,0.5\}$. Note that the turning point of $R_{\varpi}(D)$ is gained when the probability $p$ equals to the distortion $D$, which conforms to Proposition \ref{prop.binary_RD}.

\subsection{Bitrate transmission with message importance loss}
\begin{figure}[!t]
\centering
\includegraphics[width=3.75in]{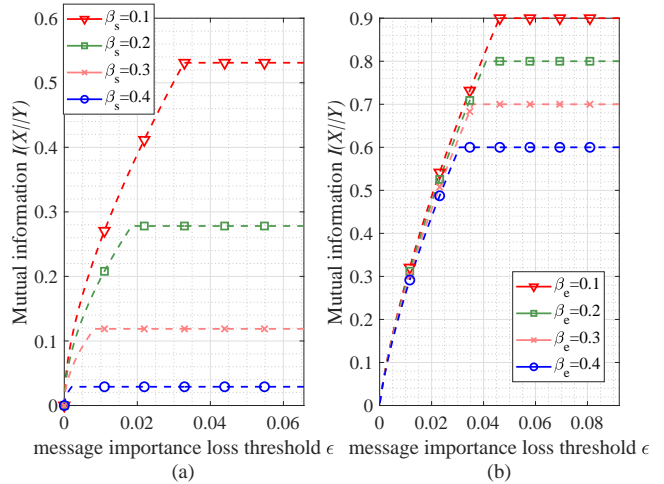}
\caption{The performance of mutual information $I(X||Y)$ constrained by the message importance loss $\epsilon$ in binary symmetric matrix (a) and erasure matrix (b) (the parameter $\varpi=0.1$). }
\label{fig_Ixy_Capacity}
\end{figure}

Fig. \ref{fig_Ixy_Capacity} shows the allowable maximum bitrate (characterized by mutual information) constrained by a message importance loss $\epsilon$ in a Bernoulli($p$) source case.
It is worth noting that there are two regions for the mutual information in the both transfer matrixes. In the first region, the mutual information is monotonically increasing with respect to the $\epsilon$, however, in the second region the mutual information is stable namely the information transmission capacity is obtained.
As for the numerical results, the turning points are obtained at $\epsilon=\{ 0.0328, 0.0185, 0.0082, 0.0021\}$ and the maximum mutual information values are $\{0.5310,0.2781,0.1187,0.0290 \}$ in the binary symmetric matrix with the corresponding parameter $\beta_s=\{0.1,0.2,0.3,0.4\}$.
While, the turning points of erasure matrix are at $\epsilon=\{ 0.0416, 0.0410, 0.0359, 0.0308\}$ in the case that $\beta_e=\{0.1,0.2,0.3,0.4\}$ with the maximum mutual information values as $\{0.9,0.8,0.7,0.6 \}$.
Consequently, the Proposition \ref{prop.symmetric_IL} and \ref{prop.error_symmetric_IL} are validated from the numerical results.


\section{Conclusion}
In this paper, we investigated an information measure i.e. MIM from the perspective of Shannon information theory.
Actually, with the help of parameter $\varpi$, the MIM has more flexibility and can be used more widely than Shannon entropy. Here, we just focused on the MIM with $ 0 \le \varpi \le 2/\max\{p(x_i)\}$ which has similarity with Shannon entropy in information compression and transmission.
In particular,
based on a system model with message importance processing, a message importance loss was presented.
This measure can characterize the information distinction before and after a message transfer process.
Furthermore, we have proposed the message importance loss capacity which can provide an upper bound for the message importance harvest in the information transmission.
Moreover, the message importance distortion function was discussed to guide the information source compression based on rare events message importance.
In addition, we exploited the message importance loss to constrain the bitrate transmission, which can add a novelty to the Shannon theory.
To give the validation for theoretical analyses, some numerical results were also presented in details.
In the future, we are looking forward to exploit the information measure theory mentioned in this paper to analyze some real databases.

\section*{Acknowledgment}
The authors appreciate for the support of the National Natural Science Foundation of China (NSFC) No. 61771283.

\ifCLASSOPTIONcaptionsoff
  \newpage
\fi

\end{document}